\documentclass[12pt]{iopart}

\usepackage{graphicx}
\usepackage{iopams}
\usepackage{cite}

\newcommand{\mrm}{\mathrm}
\usepackage{iopams}

\begin{document}

\title{Dynamics of the reading process of a quantum memory}

\author{Milrian S Mendes$^1$, Pablo L Saldanha$^{1,2}$, Jos\'e W R Tabosa,$^1$ and Daniel Felinto$^1$}
\address{$^1$Departamento de F\'isica, Universidade Federal de Pernambuco, 50670-901, Recife, PE, Brazil \\ $^2$Departamento de F\'isica, Universidade Federal de Minas Gerais, Caixa Postal 702, 30161-970, Belo Horizonte, MG, Brazil}

\ead{dfelinto@df.ufpe.br}

\begin{abstract}
The mechanism of extraction of information stored in a quantum memory is studied here in detail. We consider memories containing a single excitation of a collective atomic state, which is mapped into a single photon during the reading process. A theory is developed for the wavepacket of the extracted photon, leading to a simple analytical expression depending on the key parameters of the problem, like detuning and intensity of the read field and the number of atoms in the atomic ensemble. This theory is then compared to a large set of experimental situations and a satisfactory quantitative agreement is obtained. In this way, we are able to systematically study the saturation and spectrum of the reading process, as well as clarify the role of superradiance in the system.
\end{abstract}

\pacs{32.80.Qk, 42.50.Ct, 42.50.Ex}


\section{Introduction}

Quantum memories based on single excitations stored in atomic ensembles in free space have found a variety of applications in recent years. They have been used to store single photons from various sources~\cite{chaneliere05,eisaman05,lettner11}, as well as photonic entangled states~\cite{Choi08,Dai12}. It was also demonstrated their application for the synchronization of independent quantum systems requiring heralded preparation, like single-photon sources~\cite{matsukevich06,chen06,felinto06} or parallel pairs of entangled atomic ensembles~\cite{chou07}, has also been demonstrated, an important task for various quantum information protocols~\cite{duan01,knill01}. These capabilities, together with the ability to generate complex entangled states between various memory nodes~\cite{chen10,choi10}, make such collective atomic memories an important alternative as building blocks for more general quantum networks~\cite{kimble08}. 

For all these applications, an essential step is the final extraction of information from the atomic ensemble and its mapping into a light field, what we call here the reading process of the quantum memory. Since there is no optical cavity in the system with a mode strongly coupled to the ensemble, the efficiency of this mapping relies critically on collective enhancement of the ensemble emission into a particular photonic mode~\cite{duan01,fleischhauer00,fleischhauer00b}, besides its dependency with the usual quantities affecting the excitation of single atoms, like detuning and intensity of the exciting fields. Such collective enhancement, however, may have different natures, like the phase-matching of a nonlinear optical process~\cite{shenBook} or the superradiant emission coming from a collective entangled state~\cite{dicke54}. Moreover, the reliance on collective enhancement implies the need of a large number of atoms in the ensemble and, then, of some transparency mechanism for the extraction of a single photon from the dense medium~\cite{fleischhauer00,fleischhauer00b}.

The overall physical mechanism behind such reading process can be then quite complex, with various effects contributing to the final extraction efficiency of the stored information. Previous theoretical works on this subject have treated the problem considering the propagation of quantum fields in the medium under the conditions of Electromagnetically Induced Transparency~\cite{fleischhauer00b,lukin03,jenkins06}. This approach has successfully supported various experimental works up to now~\cite{chaneliere05,eisaman05,eisaman04,matsukevich06b}. However, for the calculation of the wavepacket of the extracted photon, it typically leads to numerical solutions, due to the complexity of solving a propagation problem in a dense medium with multiple energy levels and a strong driving field, as required to achieve both transparency and collective enhancement. In many of these previous works, there is also a focus on the combined problem of first mapping a previously free flying photon into the quantum memory and then taking it out during the reading stage~\cite{chaneliere05,eisaman05,fleischhauer00,fleischhauer00b}, with the measured and calculated efficiencies mixturing these two processes. This problem relies strongly on the detailed control of the propagation of photons in the atomic ensemble, and it is one of the main reason there has been such a large attention to propagation in the description of the reading process of such collective states. Even though this combined problem is crucial for various applications, the collective state may also be generated in situ in a heralded manner, as first suggested in~\cite{duan01} and later employed in many experiments~\cite{matsukevich06,chen06,felinto06,chou07,chen10,choi10,eisaman04,matsukevich06b}. In this later case, the efficiency of the reading process is completely decoupled from the writing process, since the reading only starts after the heralding of a succesful writing stage. The possibility of such decoupling is what enables our focus solely in the reading process in the present work.

Our approach to describe the dynamics of the reading process of such collective states, and the corresponding wavepacket of the generated photon, is quite distinct from those mentioned above, since we start by assuming that the transparency condition for the extracted photon holds in the sample. This condition is achieved by reading the ensemble with a very strong field, which ``opens'' the ensemble for the outgoing photon~\cite{fleischhauer05}. Since the medium is transparent for the photon, we may neglect its propagation in the sample and consider the photon's output state to be the superposition of the independent contributions of all atoms in the ensemble, starting in a collective state containing a single excitation. This simplifies considerably the overall theoretical analysis. We consider then the Hamiltonian evolution of the atoms of the ensemble interacting with the electromagnetic field of the reading laser and with the vacuum field. The evolution is a combination of Rabi oscillations and spontaneous decay. However, since we are dealing with an initial collective state, the usual exponential decay due to spontaneous emission presents now a superradiant enhancement towards the state that leads to the photon extraction. In the end, we arrive at a simple analytical expression for the wave function of the extracted photon, that depends on the collective behavior of the atoms and on the properties of the reading laser field. We also include in the theory the main decoherence process in our particular experimental setup: the dephasing of the atomic coherences due to the presence of an inhomogeneous magnetic field in the sample. The deduction of an analytical expression allows us to analyze in a more straightforward and intuitive way the contribution of the various effects behind the reading process. We are able, for example, to isolate the collective enhancement coming from superradiance from that coming from the phase-matching of the underlying nonlinear process. We are also able to model the saturation of the reading process and its spectrum.

The theoretical approach described in the previous paragraph was developed to model the experimental results obtained from our experimental setup to generate heralded single photons from a cloud of cold cesium atoms. This setup is described in detail in section~\ref{exper}. It implements the protocol of~\cite{duan01} adapted for the generation of heralded single photons, as first proposed in~\cite{chou04}. It also applies the four-wave-mixing configuration for photon-pair generation introduced in~\cite{balic05}, and later reproduced and improved in the works of various groups~\cite{chen06,matsukevich05,laurat06}. In section~\ref{exper} we also describe our method to measure the photon statistics of the light fields generated by the atomic ensemble, and which demonstrates their nonclassical nature. In section~\ref{sec:theory} we develop then our theoretical model for the problem and obtain the analytical expressions to compare with the experiments. The series of experimental results with the corresponding theoretical curves are presented in section~\ref{sec-exper}. The main parameters we change are the detuning and intensity of the reading field. In this way, we probe the wavepacket of the extracted photon in a wide range of conditions. We also obtain the total probability to extract the photon as a function of both intensity and detuning, revealing then the saturation behavior of the system and its spectrum. The results present an excellent quantitative agreement between theory and experiment, with the use of just a few fixed parameters to fit a large number of data. In this sense, we largely validate our simplified theoretical approach to the problem as adequate to describe such experimental system, by capturing the essential physical aspects of the problem. Finally, in section~\ref{sec:conclusion} we draw our conclusions and perspectives for future developments of this investigation.

\section{Experimental setup and methods}
\label{exper}

\begin{figure}[htb]
  \hspace{0.5cm}\includegraphics[width=17cm]{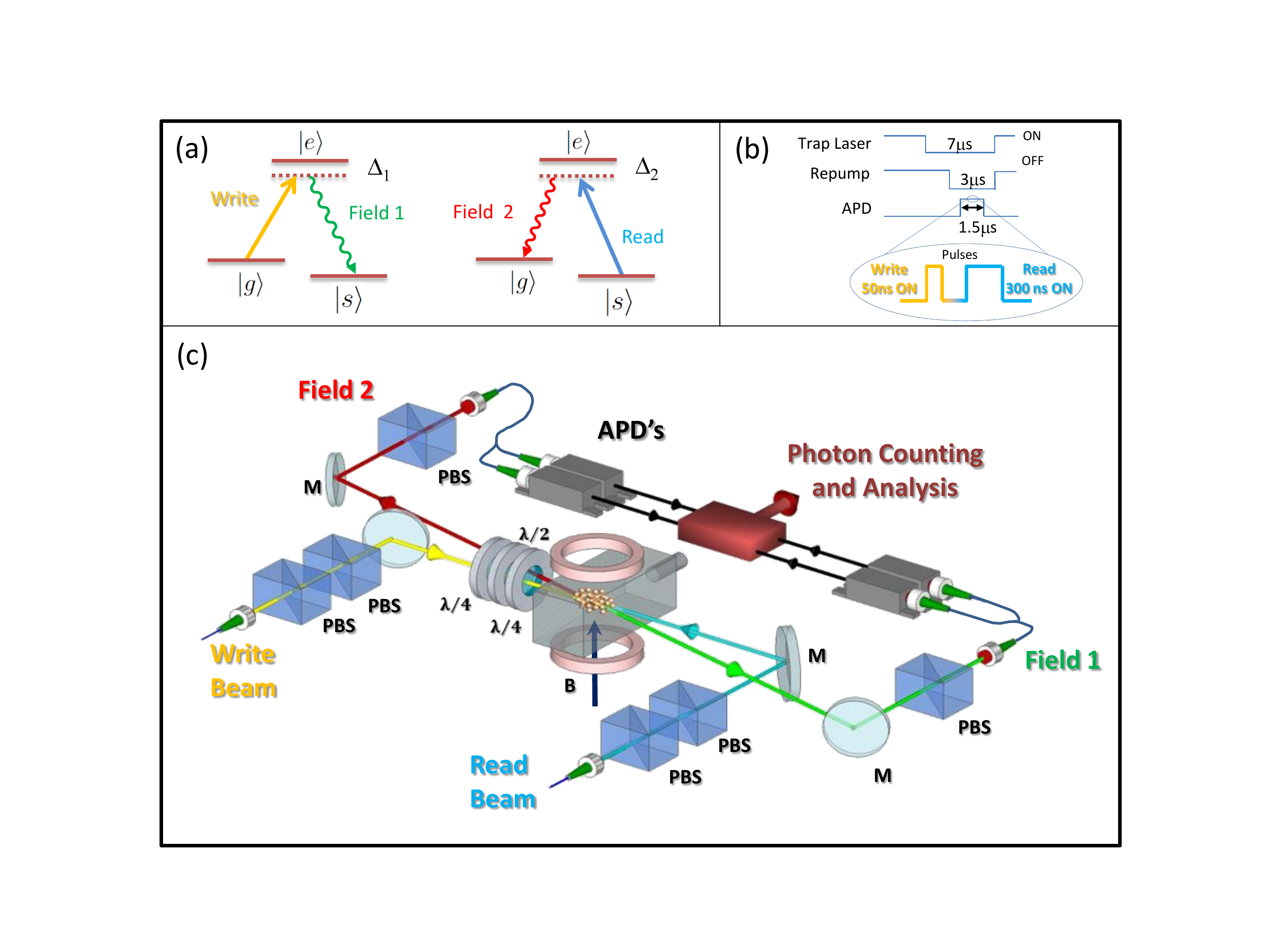}
  \vspace{-2.0cm}
  \caption{(a) $\Lambda$ configuration of levels participating in the photon-pair generation process, with indication of the transitions connected by write, photon 1, read, and photon 2 fields, respectively, and their respective detunings, $\Delta_1$ and $\Delta_2$, from the excited state. (b) Timing for the control pulses of fields and detectors participating in the experiment. (c) General description of the apparatus, see text for details. PBS stands for polarizing beam splitter, M for mirror, APD for avalanche photodetector, and B for the MOT magnetic field. Half- and quarter-wave plates are indicated by $\lambda$/2 and $\lambda$/4, respectively.}\label{fig1}
 \end{figure}

We are interested in studying the reading process of a quantum memory storing a collective atomic state encoded as a coherence grating in an ensemble of cold cesium atoms. The generation of such collective state follows the procedure originally sugested in the Duan-Lukin-Cirac-Zoller (DLCZ) protocol for quantum repeaters~\cite{duan01} and implemented for the first time in the work of \cite{kuzmich03}. In this way we consider an ensemble of three-level atoms in $\Lambda$ configuration, see figure ~\ref{fig1}(a). The two ground states $|g\rangle$ and $|s\rangle$ correspond to the hyperfine states $|6{\rm S}_{1/2}(F=4)\rangle$ and $|6{\rm S}_{1/2}(F=3)\rangle$, respectively, and the excited state $|e\rangle$ to the $|6{\rm P}_{3/2}(F^{\prime}=4)\rangle$. The cold ensemble is obtained from a magneto-optical trap (MOT), whose trapping and repumping lasers are turned off during the experiment period, see figure ~\ref{fig1}(b). The trapping laser is turned off for 7~$\mu$s and the repumping, tuned to the $|s\rangle \rightarrow |e\rangle$ transition, for just 3~$\mu$s. During 4~$\mu$s, the repumping laser is then employed to optically pump initially all atoms to $|g\rangle$. The MOT magnetic field is kept on during the whole time, which implies in very short lifetimes, tens of nanoseconds, for excitations stored in the coherence between levels $|g\rangle$ and $|s\rangle$~\cite{felinto05}. 

After the initial state preparation, a 50~ns write pulse excites the $|g\rangle \rightarrow |e\rangle$ transition, tuned $\Delta_1/(2\pi) = 20$~MHz below resonance. As a result, with small propability an atom may be transfered to state $|s\rangle$ with the simultaneous emission of a single photon in the $|e\rangle \rightarrow |s\rangle$ transition, in the optical mode we call {\it field 1}. The detection of a photon in field 1 heralds then the transfer of an atom to $|s\rangle$, but it is not known which atom in the ensemble made the transition. The ensemble is then left in a symmetrical collective state~\cite{duan01} that lives for as long as the coherence between levels $|g\rangle$ and $|s\rangle$. Since such coherence lifetime is small, the readout of the memory is performed right after the end of the write pulse. The read field is a 300~ns pulse tuned close to the $|s\rangle \rightarrow |e\rangle$ resonance, with a detuning $\Delta_2$. It maps an excitation stored in $|s\rangle$ to another photon emitted now in the $|e\rangle \rightarrow |g\rangle$ transition, in a mode we call {\it field 2}. Both write and read fields are produced inside a 1.5~$\mu$s period at which the avalanche photodetectors (APDs) for fields 1 and 2 are turned on.

For the spatial modes of the exciting fields and detected photons we employ the four-wave-mixing configuration introduced in \cite{balic05}, see figure ~\ref{fig1}(c), and later applied and perfected by various other groups~\cite{matsukevich05,laurat06,chen06}. In this way, the write field is conducted by a Polarization-Mantaining (PM) optical fiber to the experiment region, and focused in the ensemble to a diameter of 400~$\mu$m. The read field comes through a different PM fiber and arrives in the ensemble in the same transverse mode as the write field, but counterpropagating to it. For alignment, the write beam may be coupled to the read-beam fiber, and vice-versa, with about 70$\%$ coupling efficiency. The photons are emitted counterpropagating to each other forming an angle of about 1$^{\circ}$ with the direction of the write and read fields. They are also coupled to optical fibers, Single-Mode (SM) ones, whose corresponding transverse modes are focused to a diameter of 200~$\mu$m in the ensemble. An alignment laser field coming out of the field 2 fiber may be coupled with about 55$\%$ efficiency to the field 1 fiber. In such configuration field 2 is detected in the phase matched mode to field 1, with correspondingly higher probability, since the write and read fields may be approximated by plane waves due to their larger diameter in the ensemble region~\cite{laurat06}. Phase matching also requires specific combinations of polarizations between the four fields. We setup then the write and read fields with linear orthogonal polarizations. The photons 1 and 2 have also linear polarizations, with the one for field 1 being orthogonal to both write field and field 2. The polarizations of the various beams are fixed by the combination of polarizing beam splitters and waveplates shown in figure~\ref{fig1}(c).       

\vspace*{-0.6cm}
 
After the photons are coupled to their respective optical fibers, they pass through SM fiber beamsplitters and reach two independent pairs of APDs. The output of the APDs are directed then to a counting card (P7888 from Fastcomtech), which record all photodetection events for later analysis by software, with 1~ns time resolution. In this way we may compute the various single $p_i$ and joint $p_{ij}$ probabilities for all detection events in a single trial, with $i,j = 1,2$ labeled in accordance to the respective photon. From these probabilities we may calculate, for example, three important normalized correlation functions: $g_{11} = p_{11}/p_1^2$, $g_{22} = p_{22}/p_2^2$, and $g_{12} = p_{12}/p_1p_2$. The first two of these quantities measure the auto-correlations for each photon field. The third one measures cross-correlations between the two photon fields, giving the probability of generating a photon pair divided by the probability of observing an accidental coincidence event. The singular quantum nature of the correlations between fields 1 and 2 may be directly verified by the violation a Cauchy-Schwarz inequality $R = g_{12}^2/(g_{11}g_{22}) < 1$ valid for classical fields~\cite{kuzmich03,clauser74}. Since $g_{11},g_{22}$ are typically bounded by a maximum value of 2 for our system~\cite{kuzmich03}, we have that $g_{12} > 2$ also indicates purely quantum correlations between the fields. This quantity is regularly employed in our work to provide direct inference on the quantum regime of the system, since it is much easier to measure than $R$, which is plagued by the considerably smaller statistics on the measurements of $p_{11}$ and $p_{22}$. 

Finally, another important quantity is $p_\mathrm{c} = p_{12}/p_1$, which provides the conditional probability of detecting a photon 2 once a photon 1 was previously detected in the same trial. Most of the results concerning the dynamics of the reading process in the present work are related to measurements of $p_\mathrm{c}$, since it gives the probability to generate the photon 2 once the creation of the corresponding collective state is heralded by the detection in photon 1. In this sense, $p_2$ carries very different information, since it is not directly connected to the reading of a stored collective state. All the above probabilities and correlation functions may be obtained after integration throughout the whole write and read pulse durations, or over a large portion of them. Particularly, we denote by $P_c$ the integrated version of $p_c$. On the other hand, we may also compute these quantities as a function of time inside each excitation pulse. This last measurement provides then the various wavepackets for the photons, together with the corresponding details of the dynamics of the reading process in real time. Three wavepackets are of particular interest for us: $p_\mathrm{c}(t)$, $p_2(t)$, and $g_{12}(t)$.

\section{Theory}
\label{sec:theory}

Since the purpose of this work is to treat the reading process of the atomic memory, our theoretical description starts with the state of the atomic ensemble after the writing process is complete. In this case, a writing laser beam with wavevector $\bi{k}_\mrm{w}$ induces the transition $|g\rangle\rightarrow|e\rangle\rightarrow|s\rangle$ in one of the atoms of the ensemble, and a photon in a specific mode selected by an optical fiber is detected by a photon detector. The annihilation operator corresponding to the selected optical mode is given by
\begin{equation}\label{mode1}
	\hat{a}_1=\int\rmd \bi{q}_1\phi_1({\bi{q}_1})\hat{a}\left(\bi{q}_1+\sqrt{k_1^2-q_1^2}\boldsymbol{\hat{z}}\right),
\end{equation}
where $\bi{q}_1$ represents the component of $\bi{k}_1$ in the $xy$ plane and $\hat{a}(\bi{k})$ corresponds to the annihilation operator for a plane wave mode with wavevector $\bi{k}$. For simplicity, we will not consider the polarization of the optical fields here and the corresponding atomic Zeeman structure, approximating the atomic states by the three levels in Fig.~\ref{fig1}(a). The function $\phi_1({\bi{q}_1})$ defines the mode, considered to be monochromatic. A quantum state with one photon in this mode can be written as $\hat{a}_1^\dag |\mathrm{vac}\rangle$, where $\hat{a}_1^\dag$ is the creation operator for the mode and $|\mathrm{vac}\rangle$ represents the vacuum state for the electromagnetic field. We can see the writing process as the coherent scattering of one photon of the writing laser beam, which is treated as a classical field, into one photon in the mode defined by $\hat{a}_1$, together with a corresponding change on the quantum state of the atom that scattered the photon \cite{saldanha11}. Since there is a fundamental uncertainty about which atom scattered the photon, the probability amplitudes for the scattering by each atom must be coherently added. So the quantum state of the atomic ensemble at the time when the photon emitted in mode (\ref{mode1}) is detected can be computed by
\begin{equation}
    |\Psi\rangle\propto\langle \mathrm{vac}|\hat{a}_1\left[\sum_{i} E_0\rme^{\rmi\bi{k}_\mrm{w}\cdot\bi{r}_i}\int\rmd^3k'\hat{a}^\dag(\bi{k}')\rme^{-\rmi\bi{k}'\cdot\bi{r}_i}|s\rangle_i\langle g|\right]|\mathrm{vac}\rangle|g\rangle^{\otimes N},
\end{equation}
where $E_0$ represents the amplitude of the writing beam, the summation is evaluated over all atoms of the ensemble, each one at the position $\bi{r}_i$,  and $|g\rangle^{\otimes N}$ represents a state for the ensemble with all atoms in state $|g\rangle$.

While the atomic memory stores the information, decoherence processes affect the system, such that after a long time it is not possible to extract this information anymore. The principal mechanism of decoherence in our system is the Zeeman interaction of the magnetic moment of the atoms with the magnetic field of the magneto-optical trap \cite{felinto05}. The levels $|g\rangle$ and $|s\rangle$ correspond to hiperfine sublevels of the electronic ground state of cesium atoms, while the level $|e\rangle$ corresponds to an excited state. All these levels have a nonzero magnetic moment, whose component in the direction of the external magnetic field varies along the ensemble. Thus each atom, labeled by $i$, has in general  different values for the energies $\hbar\omega_{{g}i}$, $\hbar\omega_{{s}i}$, and $\hbar\omega_{{e}i}$ of levels $|g\rangle$, $|s\rangle$,  and $|e\rangle$, respectively. As the system evolves in time, the dephasing between different atoms increases, what causes a decrease of the coherence of the collective state of the atomic ensemble \cite{felinto05}.

Using the paraxial approximation $\sqrt{k_1^2-q_1^2}\approx k_1-q_1^2/(2k_1)$, if the photon detection occurs at time $t=-\tau$, the quantum state of the atomic ensemble at time $t=0$ can be written as
\begin{equation}\label{initial state}
    |\Psi(0)\rangle=\sum_{i}A_i(0)|s_i,0\rangle\rme^{\rmi(\bi{k}_\mrm{w}-k_1\boldsymbol{\hat{z}})\cdot
    \bi{r}_i},
\end{equation}
where $|s_i,0\rangle$ represents a state in which atom $i$ is in state $|s\rangle$, all the others are in state $|g\rangle$ and there are no photons in the system (with the exception, of course, of the photons of the writing beam). If atom $i$ makes the transition $|g\rangle\rightarrow|e\rangle\rightarrow|s\rangle$, from time $t=-\tau$ to $t=0$ it acquires a phase $\rme^{-i\omega_{si}\tau}$ instead of a phase $\rme^{-i\omega_{gi}\tau}$ due to the Hamiltonian evolution. So we have 
\begin{equation}\label{psi i}
	A_i(0)=c_i\rme^{-\rmi(\omega_{si}-\omega_{gi})\tau},
\end{equation}
with 
\begin{equation}\label{moder}
	c_i=\int\rmd \bi{q}_1 \phi_1({\bi{q}_1})\rme^{\rmi[-\bi{q}_1\cdot\boldsymbol{\rho}_i+z_iq_1^2/(2k_1)]}
\end{equation}
 and $\sum_i|c_i|^2=1$, where $\boldsymbol{\rho}_i$ is the component of $\bi{r}_i$ in the $xy$ plane.

\subsection{Atomic Dynamics in the Reading Process}

We consider that at time $t=0$ a reading laser beam starts to interact with the atomic ensemble, initially in the state given by (\ref{initial state}). The reading beam has amplitude $u(\bi{r},t)$, wavevector $\bi{k}_\mrm{r}$, and frequency $\omega_\mrm{r}=ck_\mrm{r}$, with the detuning for atom $i$ given by $\Delta_i=\omega_{ei}-\omega_{si}-\omega_\mrm{r}$. This beam induces the transition $|s\rangle \rightarrow|e\rangle\rightarrow|g\rangle$ on the atom in state $|s\rangle$, with the simultaneous emission of a photon whose properties depend on the collective atomic state. Treating the reading beam as a classical field and the other modes of the electromagnetic field as quantum fields initially in vacuum, the Hamiltonian that governs the time evolution of the system can be written as
\begin{equation}\label{H}
    \hat{H}=\hat{H}_0+\hat{H}_1+\hat{H}_2+\hat{H}_3
\end{equation}
with
\begin{equation}\label{H0}
    \hat{H}_0 = \sum_i\Big[\hbar\omega_{gi}|g_i\rangle\langle g_i|+\hbar\omega_{si}|s_i\rangle\langle s_i|+\hbar\omega_{ei}|e_i\rangle\langle e_i|\Big]+
  \int d^3k\;\hat{a}^\dag(\bi{k})\hat{a}(\bi{k}),
\end{equation}
\begin{equation}\label{H1}\nonumber
    \hat{H}_1 = -\sum_i\hbar g_{es}^* \hat{\sigma}_{es,i}u(\bi{r}_i,t)\rme^{\rmi[\bi{k}_r\cdot
  \bi{r}_i-\omega_rt+(\omega_{ei}-\omega_{si})t]}+\;\rm{H.\,c.},
\end{equation}
\begin{equation}\label{H2}\nonumber
  \hat{H}_2 = -\sum_i\int d^3k\; \hbar g_{eg,\bi{k}}^{*} \hat{\sigma}_{eg,i}\hat{a}(\bi{k})\rme^{\rmi[\bi{k}\cdot
  \bi{r}_i-\omega_k
  t+(\omega_{ei}-\omega_{gi})t]}+\;\rm{H.\,c.},
\end{equation}
\begin{equation}\label{H3}\nonumber
  \hat{H}_3 = -\sum_i\int d^3k\; \hbar g_{es,\bi{k}}^{*} \hat{\sigma}_{es,i}\hat{a}(\bi{k})\rme^{\rmi[\bi{k}\cdot
  \bi{r}_i-\omega_k
  t+(\omega_{ei}-\omega_{si})t]}+\;\rm{H.\,c.},
\end{equation}
where H. c. stands for the Hermitian conjugate, $\hat{\sigma}_{es,i}\equiv |e\rangle_i\langle s|$ and so on. $\hat{H}_0$ is the free Hamiltonian for the atoms of the ensemble and the electromagnetic field. $\hat{H}_1$ corresponds to the term that governs the interaction of the atoms with the incident reading beam, which induces transitions between levels $|s \rangle$ and $|e \rangle$. $g_{es}$ depends on the dipole moment of the transition. This term generates Rabi oscillations between levels $|s \rangle$ and $|e \rangle$ \cite{scully}. $\hat{H}_2$ and $\hat{H}_3$ correspond to the terms that govern the interaction of the atoms with other modes of the electromagnetic field, which are treated as quantum fields, inducing a spontaneous decay from level $|e \rangle$ to levels $|g \rangle$ or $|s \rangle$  with the emission of one photon  \cite{scully}. $g_{eg,\bi{k}}$ and $g_{es,\bi{k}}$ depend on the dipole moment of the transition and on the wavevector $\bi{k}$ of the interacting electromagnetic mode.

Considering the initial state (\ref{initial state}) for the system, the Hamiltonian evolution generates the following general state in the interaction picture:
\begin{eqnarray}\label{state}\nonumber
    |\Psi(t)\rangle=&&\sum_i \rme^{\rmi(\bi{k}_\mrm{w}-k_1\boldsymbol{\hat{z}})\cdot \bi{r}_i} \Big[
    A_i(t)|s_i,0\rangle+B_i(t)|e_i,0\rangle\Big]+\int
    \rmd^3k\;C_{\bi{k}}(t)|g,1_{\bi{k}}\rangle+\\&&+\sum_i\int
    \rmd^3k\;D_{i,\bi{k}}(t)|s_i,1_{\bi{k}}\rangle,
\end{eqnarray}
where $|e_i,0\rangle$ represents a state with atom $i$ in the state $|e\rangle$, with all other atoms in the state
$|g\rangle$ and with no photons in the system, $|g,1_{\bi{k}}\rangle$ represents a state with all atoms in the state $|g\rangle$ and one photon with wavevector $\bi{k}$ in the system. $|s_i,1_{\bi{k}}\rangle$ represents a state with atom $i$ in the state $|s\rangle$, with all others in the state $|g\rangle$ and with one photon with wavevector $\bi{k}$ in the system (considering, of course, that the reading beam is a classical field).

The time evolution of the state (\ref{state}) can be computed using the Scr\"odinger equation in the interaction picture, from which we conclude that the coefficients $A_i$, $B_i$, $C_{\bi{k}}$ and $D_{i,\bi{k}}$ obey the following set of differential equations:
\begin{equation}\label{A t}
    \dot{A}_i(t)=\rmi g_{es}u^*(\bi{r}_i,t)\rme^{\rmi[-\bi{k}_r\cdot
  \bi{r}_i-\Delta_i t]}B_i(t),
\end{equation}
\begin{eqnarray}\label{B t}\nonumber
    \dot{B}_i(t)=&&\rmi g_{es}^* u(\bi{r}_i,t)\rme^{\rmi[\bi{k}_r\cdot
  \bi{r}_i+\Delta_i t]}A_i(t)+\\\nonumber
  &+&\rmi\int \rmd^3k\; g_{eg,\bi{k}}^{*}\rme^{\rmi[(-\bi{k}_w+k_1\boldsymbol{\hat{z}}+\bi{k})\cdot
  \bi{r}_i+(\omega_{ei}-\omega_{gi}-\omega_k)t]}C_{\bi{k}}(t)+\\
  &+&\rmi\int \rmd^3k\; g_{es,\bi{k}}^{*}\rme^{\rmi[(-\bi{k}_w+k_1\boldsymbol{\hat{z}}+\bi{k})\cdot
  \bi{r}_i+(\omega_{ei}-\omega_{si}-\omega_k)t]}D_{i,\bi{k}}(t),
\end{eqnarray}
\begin{equation}\label{C t}
    \dot{C}_{\bi{k}}(t)= \rmi\sum_i g_{eg,\bi{k}}\rme^{\rmi[(\bi{k}_w-k_1\boldsymbol{\hat{z}}-\bi{k})\cdot
  \bi{r}_i+(-\omega_{ei}+\omega_{gi}+\omega_k)t]}B_i(t).
\end{equation}
\begin{equation}\label{D t}
    \dot{D}_{i,\bi{k}}(t)= \rmi g_{es,\bi{k}}\rme^{\rmi[(\bi{k}_w-k_1\boldsymbol{\hat{z}}-\bi{k})\cdot
  \bi{r}_i+(-\omega_{ei}+\omega_{si}+\omega_k)t]}B_i(t).
\end{equation}

To solve this system, we try the following form for $B_i(t)$:
\begin{equation}\label{Bi}
    B_i(t)=\beta_i(t)b_i(t)\rme^{\rmi\bi{k}_r\cdot  \bi{r}_i},
\end{equation}
such that
\begin{eqnarray}\label{b t}
  \beta_i(t)\dot{b}_i(t) &=& \rmi g_{es}^* u(\bi{r}_i,t)\rme^{\rmi\Delta_i t}A_i(t) \;\;\;\rm{and} \\\label{beta t}\nonumber
  \dot{\beta_i}(t)b_i(t) &=& \rmi\int d^3k\; \Big[g_{eg,\bi{k}}^{*}C_{\bi{k}}(t)\rme^{\rmi(\omega_{ei}-\omega_{gi}-\omega_k)t}+g_{es,\bi{k}}^{*}D_{i,\bi{k}}(t)\rme^{\rmi(\omega_{ei}-\omega_{si}-\omega_k)t}\Big]\\&&\times\rme^{\rmi(-\bi{k}_w+k_1\boldsymbol{\hat{z}}+\bi{k}-\bi{k}_r)\cdot
  \bi{r}_i}.
\end{eqnarray}
The advantage of using this form of solution for $B_i(t)$ is that now we have two sets of coupled equations. Equations (\ref{A t}) and (\ref{b t}) form a system similar to the one for the Rabi oscillations dynamics, while (\ref{C t}), (\ref{D t}) and (\ref{beta t}) form a system similar to the spontaneous decay dynamics \cite{scully}. Our method of solution will be to solve the system (\ref{C t}), (\ref{D t})  and (\ref{beta t}) and substitute the results to solve the system (\ref{A t}) and (\ref{b t}).

Let us start by the system of equations (\ref{C t}), (\ref{D t})  and (\ref{beta t}). From (\ref{C t}) and (\ref{D t}) we have
\begin{equation}\label{cc}
    C_\bi{k}(t)=\rmi\int_0^t \rmd t'\,\sum_j g_{eg,\bi{k}}\rme^{\rmi[(\bi{k}_\mrm{w}-k_1\boldsymbol{\hat{z}}-\bi{k}+\bi{k}_\mrm{r})\cdot
  \bi{r}_j+(-\omega_{ei}+\omega_{gi}+\omega_k)t']}\beta_j(t'){b_j(t')},
\end{equation}
\begin{equation}\label{dc}
    D_{i,\bi{k}}(t)=\rmi\int_0^t \rmd t'\, g_{es,\bi{k}}\rme^{\rmi[(\bi{k}_\mrm{w}-k_1\boldsymbol{\hat{z}}-\bi{k}+\bi{k}_\mrm{r})\cdot
  \bi{r}_i+(-\omega_{ei}+\omega_{si}+\omega_k)t']}\beta_i(t'){b_i(t')}.
\end{equation}
Substituting in (\ref{beta t}) we have
\begin{eqnarray}\nonumber
    &&\dot{\beta_i}(t)=-\int_0^t \rmd t'\int \rmd^3k\Bigg\{|g_{es,\bi{k}}|^2 \rme^{\rmi(\omega_{ei}-\omega_{si}-\omega_k)(t-t')}\frac{b_i(t')}{b_i(t)}\beta_i(t')+\\&&\;\;\;\;+\sum_j|g_{eg,\bi{k}}|^2\rme^{\rmi[(\bi{k}_\mrm{w}-k_1\boldsymbol{\hat{z}}-\bi{k}+\bi{k}_\mrm{r})\cdot
  (\bi{r}_j-\bi{r}_i)+(\omega_{ei}-\omega_{gi}-\omega_k)(t-t')]}\frac{b_j(t')}{b_i(t)}\beta_j(t')\Bigg\}.
\end{eqnarray}
The $\rmd^3k$ integration can be written as
\begin{equation}\label{d3k}
    \int \rmd^3k=\int_0^{2\pi}\rmd\phi\int_0^\pi
    \rmd\theta\sin(\theta)\int_0^\infty \rmd\omega_k \frac{\omega_k^2}{c^3}.
\end{equation}
Due to the term $\rme^{-i\omega_k(t-t')}$, with fast oscillations for $t\neq t'$, we may approximate
\begin{equation}
      \int_0^\infty \rmd\omega_k
      {\omega_k^2}\rme^{-\rmi\omega_k(t-t')}\approx
      2\pi\delta(t-t')\omega_k^2.
\end{equation}
So we have
\begin{eqnarray}\label{beta t2}\nonumber
    &&\dot{\beta_i}(t)=-\int_0^{2\pi}\rmd\phi\int_0^\pi \rmd\theta\sin(\theta)
    \frac{2\pi\omega_k^2(|g_{eg,\bi{k}}|^2+|g_{es,\bi{k}}|^2)}{c^3}\\ &&\times\left[ 1+\frac{|g_{eg,\bi{k}}|^2}{|g_{eg,\bi{k}}|^2+|g_{es,\bi{k}}|^2}\sum_{j\neq i} \rme^{\rmi(\bi{k}_\mrm{w}-k_1\boldsymbol{\hat{z}}-\bi{k}+\bi{k}_\mrm{r})\cdot
  (\bi{r}_j-\bi{r}_i)} \frac{b_j(t)\beta_j(t)}{b_i(t)\beta_i(t)} \right]\beta_i(t).
\end{eqnarray}
We also have $|g_{eg,\bi{k}}|^2\approx|g_{es,\bi{k}}|^2$ in our system, such that we may approximate $|g_{eg,\bi{k}}|^2/(|g_{eg,\bi{k}}|^2+|g_{es,\bi{k}}|^2)\approx1/2$ in the above equation.

According to the Weisskopf-Wigner theory for the spontaneous decay, the decay rate from the state $|e\rangle$
to the states $|g\rangle$ or $|s\rangle$ for a free atom is given by \cite{scully}
\begin{eqnarray}
    \Gamma=2\int_0^{2\pi}\rmd\phi\int_0^\pi \rmd\theta\sin(\theta)
    \frac{2\pi\omega_k^2(|g_{eg,\bi{k}}|^2+|g_{es,\bi{k}}|^2)}{c^3}=\frac{\omega_k^3(p_{eg}^2+p_{es}^2)}{3\pi\hbar\varepsilon_0 c^3},
\end{eqnarray}
since we have $|g_{eg,\bi{k}}|^2={\omega_{k}p_{eg}^2\cos^2(\theta')}/[{2(2\pi)^3\varepsilon_0\hbar}]$, where $\bi{p}_{eg}=\langle e|\hat{\bi{p}}|g\rangle$, $\hat{\bi{p}}$ is the electric dipole operator and $\theta'$ is the angle between $\bi{p}_{eg}$ and the polarization vector of the emitted photon, and similarly for $|g_{es,\bi{k}}|^2$.

If the term inside the brackets in (\ref{beta t2}) was 1, we would have the same decay rate as with one free atom. If the atoms are roughly uniformly illuminated by both write and read fields, the atomic dynamics for the optical excitation $|s\rangle \rightarrow |e\rangle$ do not vary appreciably from one atom to the other. In this way, for the evaluation of (\ref{beta t2}), we may approximate $b_j(t) \approx b_i(t)$ and $\beta_j(t) \approx \beta_i(t)$. Defining
\begin{equation}\label{chi}
    \chi_i=\left[1+\frac{3}{8\pi}\sum_{j\neq i}\int_0^{2\pi}\rmd\phi\int_0^\pi \rmd\theta\sin(\theta)\cos^2(\theta') \rme^{\rmi(\bi{k}_w-k_1\boldsymbol{\hat{z}}-\bi{k}+\bi{k}_r)\cdot
  (\bi{r}_j-\bi{r}_i)} \right],
 \end{equation}
equation~(\ref{beta t}) then becomes
\begin{equation}\label{dif exp}
    \dot{\beta_i}(t)=-\frac{\chi_i\Gamma}{2}\beta_i(t),
\end{equation}
with solution
\begin{equation}\label{beta-sol}
    \beta_i(t)=\rme^{-\chi_i\Gamma t/2}
\end{equation}
for $\beta_i(0)=1$ and $\beta_i(\infty)=0$.

If $\chi_i>1$, we have an increase of the decay rate induced by the presence of the other atoms, in a phenomenon analogous to superradiance \cite{dicke54} at the single photon level. This increase of the decay rate is not the result of stimulated emission, since only one photon is emitted by the ensemble. It is an effect that depends on the coherent distribution of the excitation through the atoms of the ensemble. In other words, this superradiance is induced by the system's entanglement. This effect is analogous to the quantum-interference-initiated superradiant emission from entangled atoms recently studied by Wiegner \textit{et al.} \cite{wiegner11}, although here we consider an ensemble of atoms in a cloud and in \cite{wiegner11} the authors considered atoms distributed in a line. However the physical mechanism behind superradiance is the same in both cases: it results from the interference of different quantum paths \cite{wiegner11}. In section \ref{sec:superradiance} we will discuss more about this superradiance effect and estimate its contribution as a function of the atomic density of the ensemble.

Having found $\beta_i(t)$, we can substitute this result in the system of equations (\ref{A t}) and (\ref{b t}), that become
\begin{eqnarray}\label{ed1b}
  \dot{A}_i(t)&=& \rmi g_{es}u^*(\bi{r}_i,t)\rme^{-\rmi\Delta_i t-\chi_i\Gamma t/2}b_i(t), \\\label{ed2b}
  \dot{b}_i(t) &=& \rmi g_{es}^* u(\bi{r}_i,t)\rme^{\rmi\Delta_i
  t+\chi_i\Gamma t/2}A_i(t).
\end{eqnarray}
Considering $u(\bi{r}_i,t)$ a real constant $u$ for $t\geq0$ and $g_{es}$ also to be real, we may eliminate $A_i$ in the above system of equations and, imposing the conditions that $b_i(0)=0$ and $A_i(0)$ is given by (\ref{psi i}), obtain
\begin{eqnarray}
b_i(t)=\rmi\,\frac{c_i\;\Omega\;\rme^{-\rmi(\omega_{si}-\omega_{gi})\tau}\;\rme^{(\chi_i\Gamma + 2\rmi\Delta_i )t/4}}{\alpha_+ +\rmi\,\alpha_-} \mathrm{sinh}\left\{\left(\frac{\alpha_+ +\rmi\,\alpha_-}{2}\right)t \right\},\label{bi-sol}
\end{eqnarray}
with $\Omega\equiv2g_{es}u$ and
\begin{equation}
 \alpha_{\pm} =  \sqrt{\sqrt{ \left( \frac{\Omega^2+\Delta_i^2}{2}-\frac{(\chi_i\Gamma)^2}{8} \right)^2+\frac{\Delta_i^2(\chi_i\Gamma)^2}{4}}\mp\left(\frac{\Omega^2+\Delta_i^2}{2}-\frac{(\chi_i\Gamma)^2}{8} \right)}.\label{alpha}
\end{equation}

Substituting (\ref{bi-sol}) and (\ref{beta-sol}) in (\ref{Bi}), we arrive at 
\begin{equation} 
B_i(t)=\rmi\,\frac{c_i\;\rme^{\rmi\bi{k}_r\cdot  \bi{r}_i}\Omega\;\rme^{-\rmi(\omega_{si}-\omega_{gi})\tau}\;\rme^{(-\chi_i\Gamma + 2\rmi\Delta_i )t/4}}{\alpha_+ +\rmi\,\alpha_-}\mathrm{sinh}\left\{\left(\frac{\alpha_+ +\rmi\,\alpha_-}{2}\right)t \right\}.\label{Bi-sol}
\end{equation}
It is important to stress that we always have $\chi_i\Gamma/2>\alpha_{+}$, such that there is always an exponential decay in the terms above. Note that the coherence between levels $|e\rangle$ and $|g\rangle$ is proportional to $B_i(t)$. Similar time dependences for the optical coherence, combining exponential decay and a hyperbolic sine function, are commonly deduced by semiclassical theories for the readout of a deterministically generated coherence grating in cold atomic ensembles~\cite{moretti08}. Such similarities are expected since the problem treated here can be understood as the single-excitation limit of the deterministic problem discussed in~\cite{moretti08}, once we restrict the analysis here to heralded events in which an excitation is stored with great certainty in the atomic ensemble. The main difference with respect to these previous treatments comes from the $\chi_i$ factor multiplying $\Gamma$, i.e., the superradiant character of the emission. As pointed out above, in the present treatment this superradiant enhancement of the branching ratio for the decay from level $|e\rangle$ to $|g\rangle$ comes from the entanglement between the atoms in the initial collective state of the ensemble, together with the indistinguishability of pathways leading the atoms back to the state $|g\rangle$ after emitting a photon in field 2. 

\subsection{Wave Function of the Photon Emitted in the Reading Process}

After the above calculation for the time evolution of the atomic state, we can now proceed on finding the mode and temporal dependence of the photon emitted in the reading process. There are two possibilities for the system dynamics that lead to completely different behaviors for the emitted photon. The first situation is the one in which a spontaneous decay from level $|e\rangle$ to level $|g\rangle$ occurs before a spontaneous decay from level $|e\rangle$ to level $|s\rangle$. In this case, according to (\ref{state}), the wavevectors decomposition of the state of the emitted photon is given by $\mathrm{Lim}_{t\rightarrow\infty}C_\bi{k}(t)$ with $C_\bi{k}(t)$ given by (\ref{cc}). We can see that the summation of the terms $\rme^{\rmi(\bi{k}_\mrm{w}-k_1\boldsymbol{\hat{z}}-\bi{k}+\bi{k}_\mrm{r})\cdot\bi{r}_i}$ in (\ref{cc}) generates the directionality of the emitted photon, since $\bi{k}_\mrm{w}\approx-\bi{k}_\mrm{r}$ in the experiments and then there is constructive interference only for $\bi{k}\approx -k_1\boldsymbol{\hat{z}}$. As we will see below, the information imprinted in the quantum memory is transferred to the photonic state in this case. 

The second situation is the one in which a spontaneous decay from level $|e\rangle$ to level $|s\rangle$ occurs before a spontaneous decay from level $|e\rangle$ to level $|g\rangle$. In this case, according to (\ref{state}), if we trace out the atomic degrees of freedom we see that the wavevectors decomposition of the density matrix of the emitted photon is given by $\sum_i\mathrm{Lim}_{t\rightarrow\infty}|D_{i,\bi{k}}(t)|^2$ with $D_{i,\bi{k}}(t)$ given by (\ref{dc}). There is no directionality on the photon emission and the information imprinted in the quantum memory is lost. Other photons can be emitted in the process, since  a transition $|s\rangle \rightarrow|e\rangle\rightarrow|g\rangle$ can be further induced in atom $i$. However, the quantum state of the atomic ensemble loses its coherence with the loss of the first photon, such that this second photon will also be emitted in a random direction and with no relation to the initial quantum state of the memory. So, if the atomic density of the ensemble is small such that there are no superradiance effects on the photon emission and $\chi_i\approx1$ for all $i$, since the decay rates from level $|e\rangle$ to levels $|g\rangle$ and $|s\rangle$ are approximately the same, there is a fundamental limit of 50\% for the efficiency of this quantum memory even if all decoherence processes and losses are perfectly eliminated, and we have strong directionality in the $|e\rangle \rightarrow |g\rangle$ emission (see section ~\ref{sec:superradiance}).

Let us  consider now the situation in which the wavevectors decomposition of the state of the emitted photon is given by $\mathrm{Lim}_{t\rightarrow\infty}C_\bi{k}(t)$ with $C_\bi{k}(t)$ given by (\ref{cc}), such that the information imprinted in the memory is transferred to the extracted photon. To find the temporal dependence of the emitted photon we must perform a Fourier transform on its frequency spectrum. Let us define the photonic mode $\Psi_2$ in terms of $\bi{q}$, the component of the photon-2 wavevector in the $xy$ plane, as we did in (\ref{mode1}) for the mode of the photon detected in the writing process. We have
\begin{equation}\label{psi f}
    \Psi_2(\bi{q};t)\propto\int \rmd\omega_k \;\rme^{-\rmi\omega_k t} \mathrm{Lim}_{t\rightarrow\infty}C_\bi{k}(t).
\end{equation}

In order to obtain an analytical expression for the wavepacket of the extracted photon, we consider two main approximations. The first is that $\chi_i$ has the same value for all atoms in the ensemble, i.e., $\chi_i \approx \chi$. The second is that the dislocation of the energy levels caused by the local magnetic fields is small when compared to $\Gamma$. In this way, if $\omega_e$, $\omega_g$, and $\omega_s$ are the unperturbed values of the respective transition frequencies, we may write $\omega_{ei} \approx \omega_e$, $\omega_{gi} \approx \omega_g + \delta_{gi}$, and $\omega_{si} \approx \omega_s + \delta_{si}$, with $\delta_{gi},\delta_{si} << \Gamma$. We neglect then any dislocation of the excited state, and may approximate $\Delta_i \approx \Delta = \omega_e - \omega_s - \omega_r$ inside the coeficients $\alpha_{\pm}$. Equation~(\ref{Bi-sol}) can then be written as
\begin{equation}
B_i(t) = c_i\;\rme^{\rmi\bi{k}_r\cdot  \bi{r}_i}\;\rme^{-\rmi(\delta_{si}-\delta_{gi})\tau}\;\rme^{-\rmi\delta_{si}t/2}\; \rme^{i \Delta t} B(t) \;,
\end{equation}
with 
\begin{equation} 
B(t)=\rmi\,\frac{\Omega\;\rme^{-\rmi(\omega_{s}-\omega_{g})\tau}\;\rme^{-\chi\Gamma t/4}\;\rme^{-\rmi\Delta t/2}}{\alpha_+ +\rmi\,\alpha_-}\mathrm{sinh}\left\{\left(\frac{\alpha_+ +\rmi\,\alpha_-}{2}\right)t \right\}\label{Bt}
\end{equation}
having the same value for all atoms in the ensemble. With these approximations, substituting (\ref{cc}) in
(\ref{psi f}) using (\ref{bi-sol}), (\ref{beta-sol}) and (\ref{moder}), we obtain
\begin{eqnarray}\label{psi sum}\nonumber
    \Psi_2(\bi{q};t)\propto &&\sum_i \int \rmd \bi{q}_1 \phi_1(\bi{q}_1) g_{eg,\bi{k}}\rme^{-\rmi[(\bi{q}_1+\bi{q})\cdot\boldsymbol{\rho}_i+(k_{1z}+k_z)z]}\rme^{-\rmi(\delta_{1i}\tau+\delta_{2i}t)}\\&&\times B(t)\rme^{-\rmi\omega t},
\end{eqnarray}
since $k_{1z}\approx k_1-q_1^2/(2k_1)$ and $\bi{k}_\mrm{w}\approx-\bi{k}_\mrm{r}$, with  $\delta_{1i}\equiv\delta_{si}-\delta_{gi}$, $\delta_{2i}\equiv(\delta_{si}/2) - \delta_{gi}$, and the central frequency of the emitted photon being $\omega=\omega_e-\omega_g-\Delta$~\cite{deOliveira12}. For simplicity, we will assume $\delta_{1i} \approx \delta_{2i} \approx \delta_i$, to reduce the number of variables and since the difference between $\delta_{1i}$ and $\delta_{2i}$ will be equivalent to slightly different values of $\tau$. 

As previously discussed, different atoms have different energy levels due to the Zeeman interaction with the magnetic field of the trap. Since the mode defined by (\ref{mode1}) in general has a small width in the $xy$ plane, the variation of the magnetic field in the $z$ direction is the principal cause of decoherence in the system. The field around the $z$ axis can be approximated by $\bi{B}\approx bz\boldsymbol{\hat{z}}$, with a linear dependence on the $z$ position. The projection of the magnetic moment of the hiperfine states in the $z$ direction is not controlled, so for each set of atoms with quantum number $m_F$ we will have a different dependence of the energy levels with $z$. To simplify the calculations, we will consider an effective interaction taking the average of the Zeeman splittings, such that $\delta_i\approx \xi z_i$, $\xi$ being a constant that defines the decoherence of the system.

Let us substitute the summation in (\ref{psi sum}) by spatial integrals that contain the atomic density $ \rho(\bi{r})\propto\rme^{-z^2/(2L^2)}$ of the ensemble, where $L$ is the width of the ensemble in the $z$ direction. We consider an uniform density in the $xy$ plane because the distribution of the excitation in the ensemble given by $|c_i|^2$ from (\ref{moder}) has a width much smaller than the width of the atomic ensemble in this plane. We then obtain
\begin{equation}\label{psi}
    \Psi_2(\bi{q};t)\propto \phi_1(-\bi{q})\,\rme^{-L^2(k_{1z}+k_z)^2/2}\,\rme^{-\xi^2L^2(t+\tau)^2/2} B(t)\rme^{-\rmi\omega t}.
\end{equation}
This expression demonstrates that the second photon comes, as expected, in the conjugate mode to field 1, with $k_z \approx -k_{1z}$ and the corresponding $\phi_1$ as transversal mode. It also explicitly relates the $\xi$ parameter coming from the MOT magnetic field to a decay rate $\gamma = \xi L$. The conditional probability $p_\mathrm{c}(t)$ to detect the second photon at time $t$ once the first photon was detected is then given by
\begin{equation}
p_\mathrm{c}(t) = F \,\rme^{-\gamma^2(t+\tau)^2} |B(t)|^2 \,, \label{pc-theory}
\end{equation}
with $F$ a proportionality constant and $B(t)$ given by (\ref{Bt}). Another important quantity is the total conditional probability $P_\mathrm{c}$:
\begin{equation}
P_\mathrm{c} = \int_0^{\infty} p_\mathrm{c}(t)dt \,, \label{Pc-theory}
\end{equation}
which gives the probability to extract the photon during the whole reading process. In section ~\ref{sec-exper}, we compare the predictions of~(\ref{pc-theory}) and~(\ref{Pc-theory}) to a series of experimental results.

\subsection{Superradiance induced by the system entanglement}\label{sec:superradiance}

Before we proceed with the comparison of our experimental results to the above theory, we are going to deduce an expression to estimate the role of superradiance in typical experimental conditions. For that we compute $\chi_i$ from (\ref{chi}) as a function of the atomic density of the ensemble under some approximations. First of all, let us disregard the dipole radiation pattern of the atomic emission, considering instead an uniform emission in all directions, and substitute again the summation by integrals over the atomic density. With these considerations, and since $\bi{k}_\mrm{w}\approx-\bi{k}_\mrm{r}$ and $k_1\approx k$, $\chi_i$ may be written as
\begin{eqnarray}\label{chi aux}\nonumber
	\chi_i=1+&&\frac{1}{2\pi k^2}\int_{-k}^{k}\rmd q_x \int_{-\sqrt{k^2-q_x^2}}^{\sqrt{k^2-q_x^2}} \rmd q_y \int \rmd^3r\rho(\bi{r})\rme^{-\rmi[q_x(x-x_i)+q_y(y-y_i)]}\\&&\times\left\{\rme^{-\rmi(k-\sqrt{k^2-q_x^2-q_y^2})(z-z_i)}+\rme^{-\rmi(k+\sqrt{k^2-q_x^2-q_y^2})(z-z_i)}\right\}.
\end{eqnarray}
To evaluate the volume integral, we will consider that $\rho$ is given by 
\begin{equation}\label{sum-int}
	\rho(\bi{r})=\frac{N}{(2\pi)^{3/2}W^2L}\rme^{-(x^2+y^2)/(2W^2)}\rme^{-z^2/(2L^2)},
\end{equation}
where $W$ is the waist of the mode of the detected photon (considered to be Gaussian), $L$ is the width of the ensemble in the $z$ direction and $N$ is the total number of atoms in this region. Only the atoms that are in the region of the mode of the detected photon can store the excitation, so only these atoms are considered to compute $\chi_i$. Evaluating the volume integral we obtain
\begin{eqnarray}\nonumber
	\chi_i=1+&&\frac{N}{2\pi k^2}\int_{-k}^{k}\rmd q_x \int_{-\sqrt{k^2-q_x^2}}^{\sqrt{k^2-q_x^2}} \rmd q_y \exp\left[\frac{-W^2(q_x^2+q_y^2)}{2}\right]\rme^{-\rmi(q_xx_i+q_yy_i)}\\\nonumber
	&&\times\Bigg\{ \exp\left[ \frac{-L^2(k-\sqrt{k^2-q_x^2-q_y^2})^2}{2} \right]\rme^{-\rmi(k-\sqrt{k^2-q_x^2-q_y^2})z_i} +\\
	&&\;\;\;\;+\exp\left[ \frac{-L^2(k+\sqrt{k^2-q_x^2-q_y^2})^2}{2} \right]\rme^{-\rmi(k+\sqrt{k^2-q_x^2-q_y^2})z_i}\Bigg\}.
\end{eqnarray}

For the usual atomic ensembles used for performing quantum memories, typical values for the quantities $k$, $W$ and $L$ are $k\approx10^{7}$m$^{-1}$, $W\approx10^{-4}$m and $L\approx10^{-3}$m. The function $\exp[-W^2(q_x^2+q_y^2)/2]$ has width $1/W<<k$ in $q_x$ and $q_y$, such that the integrals in $q_x$ and $q_y$ can be extended from $-\infty$ to $+\infty$ and we may approximate $k-\sqrt{k^2-q_x^2-q_y^2}\approx (q_x^2+q_y^2)/(2k)$. On this way, the function $\exp\{-L^2[(q_x^2+q_y^2)/(2k)]^2/2\}$ has width $\sqrt{2}k/L$ in $q_x^2+q_y^2$, while the function $\exp[-W^2(q_x^2+q_y^2)/2]$ has width $1/W^2\ll\sqrt{2}k/L$ in $q_x^2+q_y^2$, so the first of these functions can be considered unity for the evaluation of the integrals in $q_x$ and $q_y$. We can also approximate $\exp[-L^2k^2]\approx0$. Evaluating the above integrals under these approximations and disregarding terms with $L/(W^2k)$ in relation to $1$, we obtain
\begin{equation}
	\chi_i=1+\frac{N}{W^2k^2}\exp\left[\frac{-(x_i^2+y_i^2)}{2W^2}\right].
\end{equation}
Considering the distribution of the excitation in the ensemble following $\rho/N$ from (\ref{sum-int}), the average $\chi$ is given by
\begin{equation}\label{chi ap}
	\chi=1+\frac{N}{2W^2k^2}.
\end{equation}
When $\chi>1$, according to (\ref{beta-sol}) the spontaneous decay rate from level $|e\rangle$ in the reading process increases, a characteristic feature of superradiance~\cite{dicke54}.

It is worth mentioning that the directionality on the photon emission in the reading process, as predicted by (\ref{psi}), depends only on the extension of the atomic ensemble over distances large compared to $\lambda$. Once this condition is fulfilled, the directionality in our process grows proportionally to $N^2$, a very well known effect in such four-wave mixing systems. In this way, it is possible to achieve strong directionality without superradiance, i.e., with $\chi \approx 1$. In this situation, as we discussed in the previous section, there is a fundamental limit of 50\% on the efficiency of the quantum memory due to the spontaneous decay from level $|e\rangle$ to level $|s\rangle$. Of course, since this limit comes from the branching ratio of the various decay channels of the excited state, it should decrease in the actual experiment with real atoms and their whole Zeeman structure. For typical experiments with alkali atoms, like cesium and rubidium, each excited state should have about six decay channels, corresponding to transitions with a variation $\Delta m_F = 0$ or $\pm1$ of the magnetic quantum number and to one of the two hyperfine ground states.

On the other hand, when $\chi>1$ this efficiency may increase, since the decay rate from level $|e\rangle$ to level $|g\rangle$ increases in relation to the decay rate from level $|e\rangle$ to level $|s\rangle$ due to the superradiance effect, as we show below. From (\ref{D t}) we have 
\begin{equation}	\int_0^{2\pi}\rmd\phi\int_0^\pi\rmd\theta\sin(\theta)\sum_i|\dot{D}_{i,\bi{k}}|^2=\sum_i\int_0^{2\pi}\rmd\phi\int_0^\pi\rmd\theta\sin(\theta)|g_{es,\bi{k}}|^2|B_i|^2,
\end{equation}
while from (\ref{C t}) we have
\begin{eqnarray}\nonumber	  &&\int_0^{2\pi}\rmd\phi\int_0^\pi\rmd\theta\sin(\theta)|\dot{C}_{\bi{k}}|^2=\sum_i\int_0^{2\pi}\rmd\phi\int_0^\pi\rmd\theta\sin(\theta)|g_{eg,\bi{k}}|^2|B_i|^2\\&&\;\;\;\;\;\;\;\times\left[1+\sum_{j\neq i} \rme^{\rmi(\bi{k}_\mrm{w}-k_1\boldsymbol{\hat{z}}-\bi{k}+\bi{k}_\mrm{r})\cdot
  (\bi{r}_j-\bi{r}_i)} \frac{\beta_jb_j}{\beta_ib_i}\right].
\end{eqnarray} 
Comparing the above equation with (\ref{beta t2}) and (\ref{dif exp}) and approximating $\beta_j\approx\beta_i$, $b_j\approx b_i$, and $|g_{es,\bi{k}}|^2\approx|g_{eg,\bi{k}}|^2$, we obtain 
\begin{equation}	
\int_0^{2\pi}\rmd\phi\int_0^\pi\rmd\theta\sin(\theta)|\dot{C}_{\bi{k}}|^2=(2\chi-1)\int_0^{2\pi}\rmd\phi\int_0^\pi\rmd\theta\sin(\theta)\sum_i|\dot{D}_{i,\bi{k}}|^2,
\end{equation}
such that the ratio between the decay rate from state $|e\rangle$ to state $|g\rangle$ and to state $|s\rangle$ in the reading process is $2\chi-1$. So, the higher the value of $\chi$, the more efficient is the overall quantum memory readout.

\section{Results}
\label{sec-exper}

In the previous section we introduced an analytical theory to describe the reading process of our quantum memory that assumes a series of reasonable approximations, while still capturing the essential physical aspects of the problem. In the following we provide a systematic study of the reading process varying two of its main parameters: detuning and intensity of the read beam. We measure then the wavepackets and overall extraction probability of photon 2 for a large number of parameters and compare them with our theory. We obtain a quantitative agreement between theory and experiment that validates our theoretical approach to the problem to a great extent.

The theory developed so far provides directly the wavepacket of the extracted photon, describing then the dynamics of the reading process. It is not a theory designed for the verification of the purely quantum nature of the measured field, i.e., we do not calculate the photon statistics of the extracted optical fields. In order to verify its quantum nature, we employ standard quantum optical measurements for the correlation functions of the combined system of fields 1 and 2~\cite{kuzmich03,clauser74}. Such correlation measurements are provided in the following sections in addition to the measurements for direct comparison the theory. Basically, all presented results were obtained well in the purely quantum regime for the memory. 

Sections~\ref{wpct} and~\ref{sat} present results obtained at the same day under the same experimental conditions. In this way, we are able to fit all experimental results with the same set of theoretical parameters in (\ref{pc-theory}). For the decoherence parameter, we employed $\gamma / (2\pi) = 1.55$~MHz, consistent with typical numbers obtained from other single-photon-generation measurements from magneto-optical traps without turning off the trapping magnetic field~\cite{felinto05}. For the saturation intensity, we found $I_\mathrm{s} = 12$~mW/cm$^2$, as defined through the relation $(\Omega/\Gamma)^2 = I_\mathrm{r}/(2I_\mathrm{s})$ between the Rabi frequency and the intensity $I_\mathrm{r}$ of the read beam in the experiment. The superradiant, coorperativity parameter that best fit our data was found to be $\chi = 2.7$, indicating then that we are already in the regime $\chi > 1$ in which superradiance have a significant role. Finally, for the proportionality parameter we obtained $F = 4.1$. The results in section ~\ref{spec} were obtained a couple of days later under roughly the same conditions. The only change in the fitting parameters was the $F=4.8$ we employed for this set of data, reflecting a better fiber coupling on this day. It is important to keep in mind, however, that we are here comparing a simplified theory considering only three atomic levels with experiments involving a more complex level structure. The theoretical parameters reflect then just effective values for these quantities under the specific approximate model.

In order to determine a $\chi > 1$ experimentally, one approach could be to measure the overall conditional probability $P_\mathrm{c}$, compensate for all known losses, and check if the corrected extraction probability is larger than the limit one would expect without superradiance, as discussed in the previous section. For a three-level atom, if this extraction probability is then higher than 50$\%$, we could apply the relation deduced in section ~\ref{sec:superradiance} between $\chi$ and the branching ratio to obtain an estimation for its value. This approach, however, has the drawback of requiring a prior knowledge of all other loss mechanisms in the experimental setup. In our setup, for example, the MOT magnetic field induce large losses in the photon extraction, due to decoherence, which are difficult to accurately quantify independently. Even if the MOT magnetic field is off, however, unknown extra losses could easily decrease any corrected extraction probability, leading to an underestimation of $\chi$. Our approach here goes in a different direction. We obtain $\chi$ from fittings of the experimental data for various intensities and detunings of the read field, testing the effect of an effective change in $\Gamma$ over the saturation and lineshape of the reading process. The quality of the final fitting is then crucial to guarantee the significance of the value found for $\chi$. The drawbacks of this approach are its dependence on a particular model for the reading process and the correspondingly indirect determination of $\chi$. A better method would be to combine, in the future, the two approaches, seeking their convergence and using the indirect measurement to corroborate the losses estimation employed in direct measurements of $\chi$.        

\subsection{Wavepackets}
\label{wpct}

\begin{figure}[htb]
  \hspace{3.0cm}\includegraphics[width=10cm]{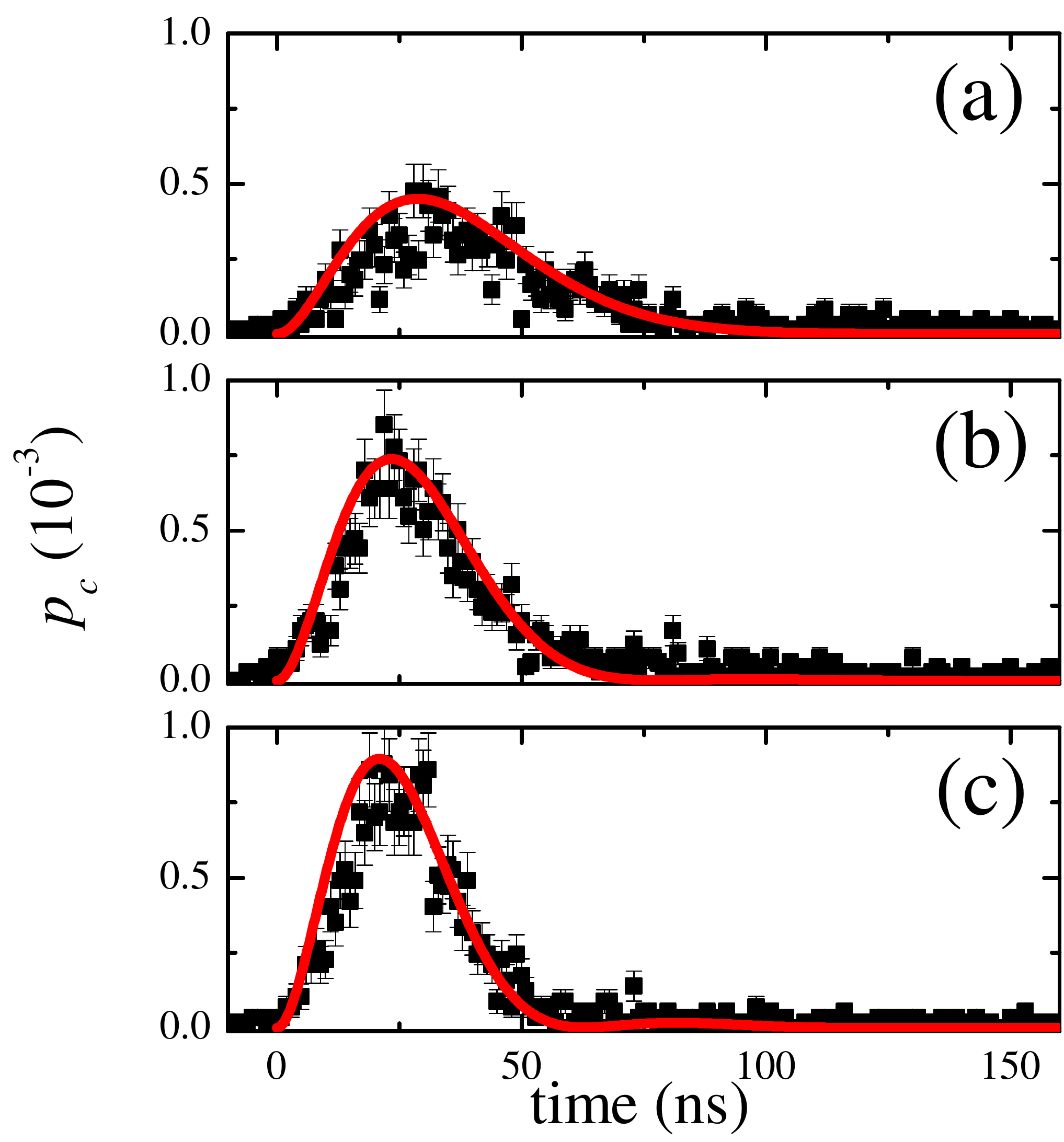}
  \vspace{-0.0cm}
  \caption{Conditional probability $p_\mathrm{c}$ for detecting a photon 2, once a photon was previously detected in field 1, as a function of time for various intensities of the read beam. The black squares are the experimental results for (a) $I_\mathrm{r} = 32$~mW/cm$^2$, (b) $I_\mathrm{r} = 68$~mW/cm$^2$, and (c) $I_\mathrm{r} = 95$~mW/cm$^2$. The detuning of the read laser is $\Delta / (2\pi) = 1.7$~MHz. The red curves are the corresponding theoretical predictions from (\ref{pc-theory}), considering $I_\mathrm{s} = 12$~mW/cm$^2$, $\gamma / (2\pi)= 1.55$~MHz, $\chi = 2.7$, and $F = 4.1$.}\label{fig2}
\end{figure}

\begin{figure}[htb]
  \hspace{3.0cm}\includegraphics[width=10cm]{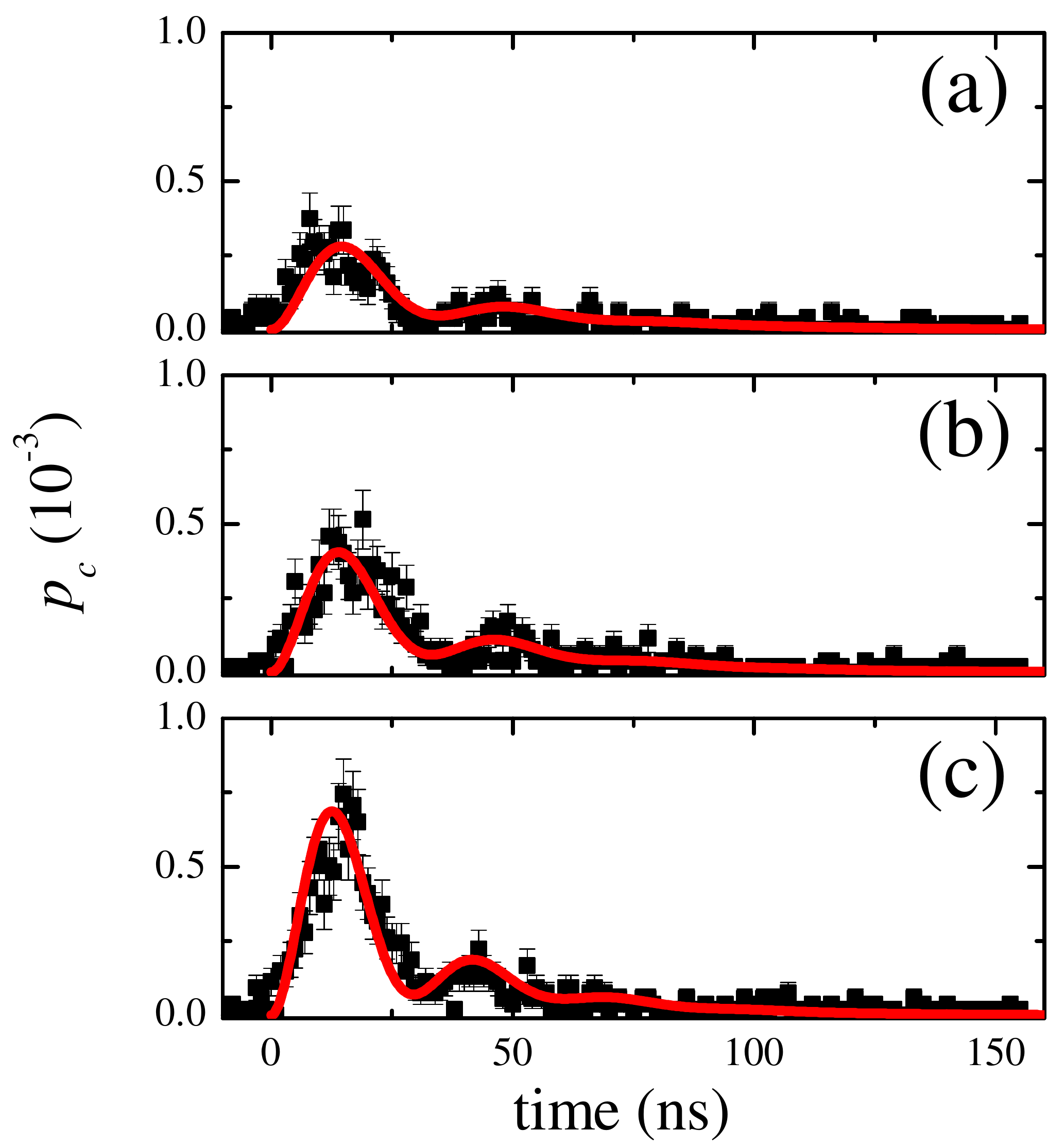}
  \vspace{-0.0cm}
  \caption{Conditional probability $p_\mathrm{c}$ for detecting a photon 2, once a photon was previously detected in field 1, as a function of time for various intensities of the read beam. The black squares are the experimental results for (a) $I_\mathrm{r} = 52$~mW/cm$^2$, (b) $I_\mathrm{r} = 80$~mW/cm$^2$, and (c) $I_\mathrm{r} = 160$~mW/cm$^2$. The detuning of the read laser is $\Delta/ (2\pi)= 25.7$~MHz. The red curves are the corresponding theoretical predictions from (\ref{pc-theory}), considering the same fitting parameters of figure ~\ref{fig2}.}\label{fig3}
\end{figure}

The black squares in Figs.~\ref{fig2} and~\ref{fig3} are experimental results for the conditional probability $p_\mathrm{c}$ as a funtion of time, with the time origin in the moment the read field is turned on. The data is presented with points separated by the acquisition-board maximum resolution, 1~ns. In figure ~\ref{fig2} we plot the results for the read detuning $\Delta / (2\pi)= 1.7$~MHz and three different read intensities: $I_\mathrm{r} = 32$~mW/cm$^2$, $68$~mW/cm$^2$, and $95$~mW/cm$^2$. Figure~\ref{fig3} plots the results for $\Delta / (2\pi)= 25.7$~MHz and the intensities $I_\mathrm{r} = 52$~mW/cm$^2$, $80$~mW/cm$^2$, and $160$~mW/cm$^2$, respectively. The error bars represent the statistical uncertainty for the counts in each time beam. The strong read fields are crucial for these measurements to guarantee the transparency of the medium for the extracted photon, as well as a fast readout in face of our very short coherence times. 
Other important experimental parameters are the optical depth $OD \approx 5$ of the atomic ensemble and the probability $p_1 = 0.0036 \pm 0.0004$ for detecting a photon in field 1. OD was determined from the absorption of a linearly polarized short pulse, 0.5~$\mu$s, resonant with the $|g\rangle \rightarrow |e\rangle$ transition, and propagating through the ensemble in the transversal mode of the write, read fields. The $\Delta = 0$ position was determined by an independent measurement of the absorption lineshape of the read field when tuned around the $|s\rangle \rightarrow |e\rangle$ transition.
 
\vspace*{-0.0cm}

Wavepacket measurements such as the ones in Figs.~\ref{fig2} and~\ref{fig3} provide the most direct experimental observations of the dynamics of the reading process, with its rise and decay times and eventual oscillatory behavior. The corresponding theoretical curves obtained from (\ref{pc-theory}) are given by the red curves in each figure. We obtain then a satisfactory quantitative agreement for both the pulse shapes and extraction efficiencies, employing the single set of fitting parameters provided above for all curves.

Figure~\ref{fig4}, on the other hand, plot the normalized cross-correlation function $g_{12} = p_\mathrm{c}(t)/p_2(t)$ as a function of time for $I_\mathrm{r} = 95$~mW/cm$^2$ and two detunings, $\Delta/(2\pi) = 1.7$~MHz and 25.7~MHz. The resolution in time was reduced to 3~ns to allow for a better statistics at each point. The time behavior of $g_{12}$ and $p_\mathrm{c}$ have pronounced differences, most strickingly the saturation of the maxima at $g_{12} \approx 20$ for both curves. As anticipated, the theory of section ~\ref{sec:theory} do not model such experimental data, since we do not calculate the unconditioned states of light originating $p_2(t)$. However, as discussed in section ~\ref{exper}, the condition $g_{12}>2$ is a strong indication of the purely quantum nature for the correlations between fields 1,2. The plots in figure ~\ref{fig4} provide the time behavior of such nonclassical correlations, which can be used to select optimum time windows to perform quantum information protocols~\cite{polyakov04}.  
  
\vspace*{-0.0cm}

\subsection{Saturation}
\label{sat}

\begin{figure}[htb]
  \hspace{3.0cm}\includegraphics[width=10cm]{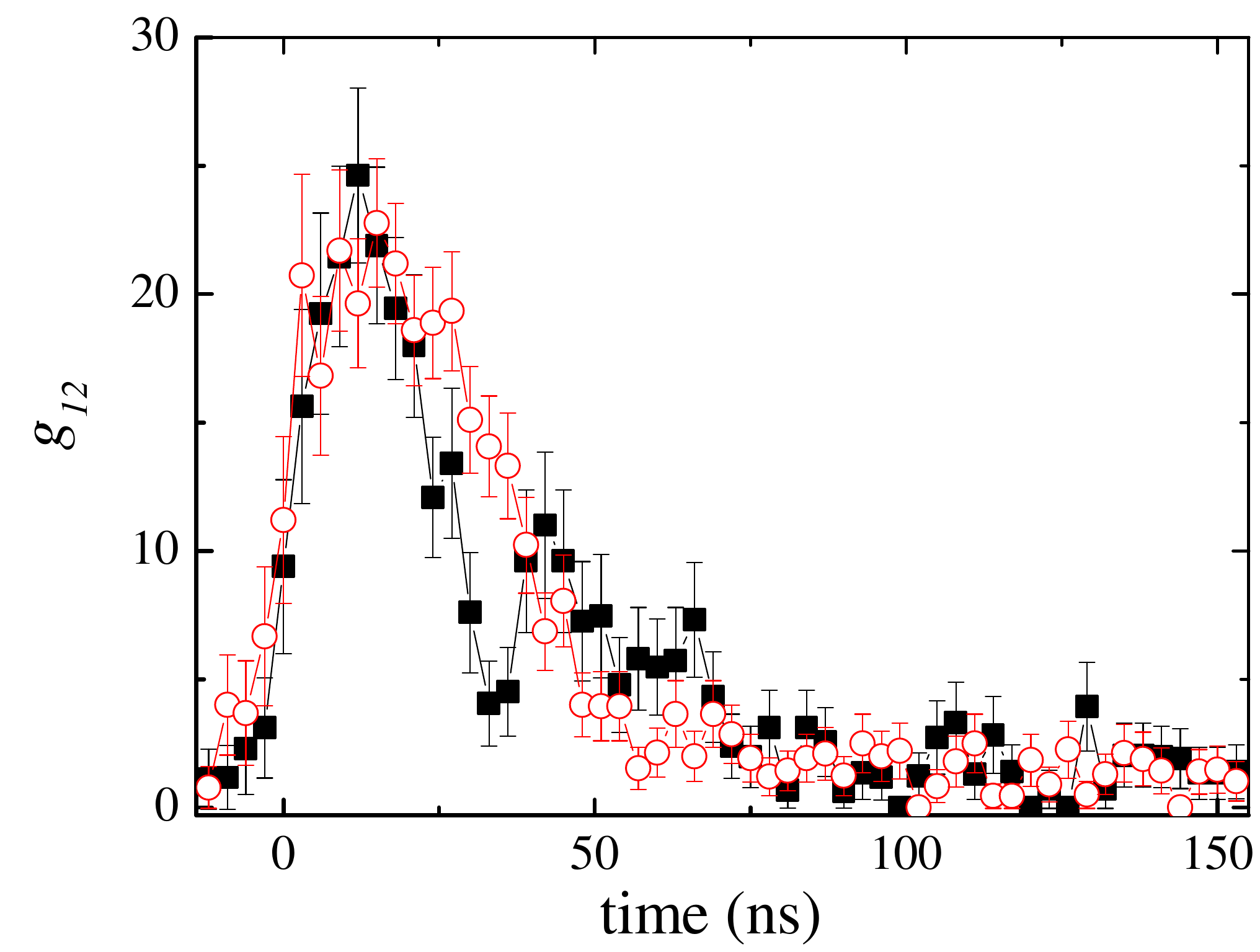}
  \vspace{-0.0cm}
  \caption{Normalized cross-correlation function $g_{12}$ as a function of time for $\Delta / (2\pi)= 25.7$~MHz (black squares) and 1.7~MHz (open red circles), respectively. The intensity $I_\mathrm{r} = 95$~mW/cm$^2$ was the same for both detunings. The lines are just guides to the eyes.}\label{fig4}
\end{figure}

\begin{figure}[htb]
  \hspace{3.0cm}\includegraphics[width=10cm]{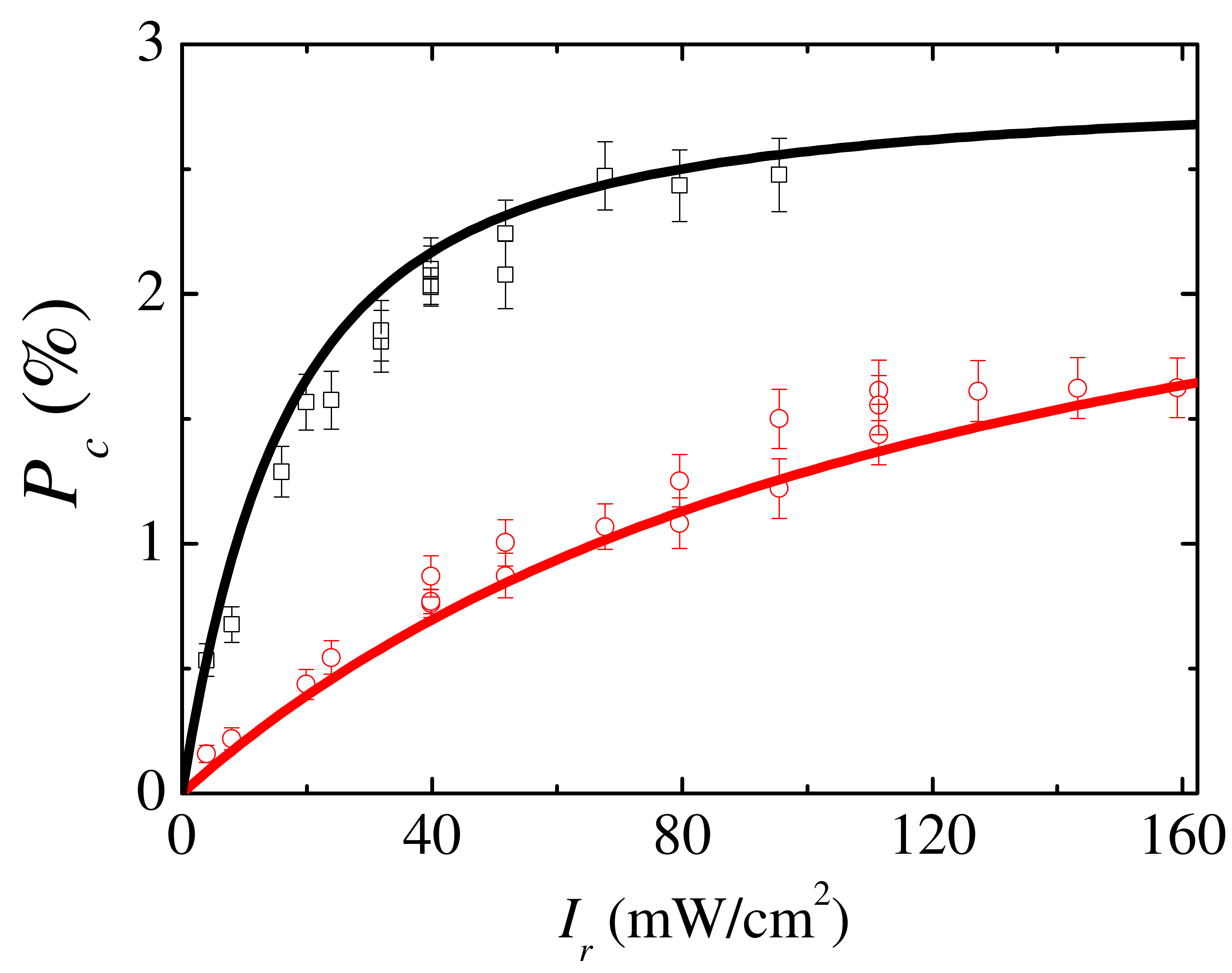}
  \vspace{-0.0cm}
  \caption{Total conditional probability $P_\mathrm{c}$ as a function of read-field intensity $I_\mathrm{r}$ for two detunings: $\Delta / (2\pi)= 1.7$~MHz (black squares) and 25.7~MHz (red circles). The solid lines are the corresponding theoretical results obtained from (\ref{Pc-theory}) with the same parameters of figure ~\ref{fig2}.}\label{fig5}
\end{figure}

In order to characterize the saturation of the total extraction probability $P_\mathrm{c}$ with respect to the read-field intensity $I_\mathrm{r}$, we integrate the wavepackets given by $p_\mathrm{c}(t)$ over a large time window, the total 160~ns of Figs.~\ref{fig2}-\ref{fig4}. The results for $P_\mathrm{c}$ as a function of $I_\mathrm{r}$ for two different detunings, $\Delta / (2\pi)= 1.7$~MHz and 25.7~MHz, are shown in figure ~\ref{fig5}. The corresponding theoretical curves are given by the solid lines fitting the experimental points. Some of the experimental points were measured more than once to provide an estimate for the long term fluctuations in our system, which may result in dispersion of the experimental points on the top of the usual statistical uncertainties. 

Such integrated results provide a broader picture of the dependence of the readout process with its main experimental parameters. In figure ~\ref{fig5} we note, for example, that the read intensity for which $P_\mathrm{c}$ saturates and the maximum value of $P_\mathrm{c}$ depend both strongly on $\Delta$. Such saturation for $P_\mathrm{c}$ also occurs at much higher values of $I_\mathrm{r}$ than one would expect from the saturation intensity $I_\mathrm{s} = 12$~mW/cm$^2$ obtained from our global fitting. We understand this effect as coming from our short coherence times due to the MOT magnetic fields. In order to extract the photon, we need to extract it fast when compared to this coherence time, and this requires higher intensities. Higher intensities are also required to guarantee the transparency of the medium to the extracted photon, which will be more cleraly revealed in section ~\ref{spec}. 

\vspace*{-0.0cm}
 
\begin{figure}[htb]
  \hspace{3.0cm}\includegraphics[width=10cm]{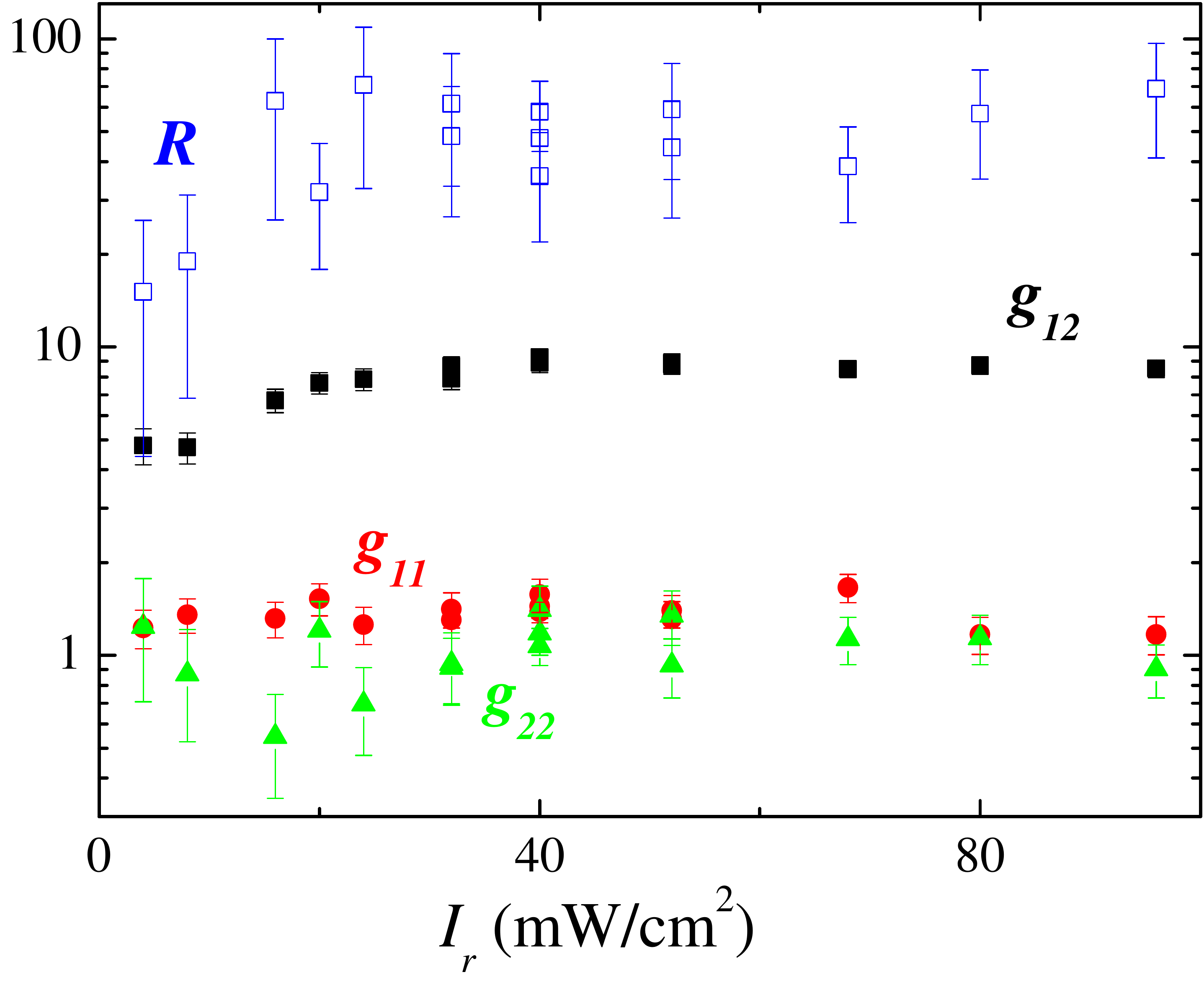}
  \vspace{-0.0cm}
  \caption{Experimental quantities characterizing the correlation between fields 1 and 2 as a function of the read intensity $I_\mathrm{r}$, for $\Delta / (2\pi)= 1.7$~MHz. The black, solid squares are the results for the normalized cross-correlation function between field 1 and 2, $g_{12}$. The red, solid circles (green, solid triangles) are the results for the normalized auto-correlation function of field 1 (2), $g_{11}$ ($g_{22}$). The blue, open squares are the results for function $R$, which indicates the nonclassical nature of the correlations if $R>1$.}\label{fig6}
\end{figure}

Figure~\ref{fig6} provide then results demonstranting directly the violation of the Cauchy-Schwarz inequality $R = g_{12}^2/(g_{11}g_{22}) < 1$ valid for classical fields, where $g_{ij}$ (with $i,j = 1,2$) are the various correlations functions defined in section ~\ref{exper} between fields 1 and 2. The quantities in figure ~\ref{fig6} are calculated for the points in figure ~\ref{fig5} with $\Delta / (2\pi)= 1.7$~MHz. We also employed here the 160~ns time window of the previous figure. This is crucial to improve the statistics of the measured quantities. Even though, we still have large statistical uncertainties for the determination of $g_{22}$, and also a large susceptibility of both $g_{11}$ and $g_{22}$ for our long term experimental fluctuations. For their determination, these quantities require measurements of the two-photon components in fields 1 and 2, which are quite low once we enter well into the single-photon regime, as indicated by $g_{12} \approx 9$~\cite{laurat06}. This leads to the large fluctuations and error bars in $R$. We clearly observe, however, $R>>1$ for all measured values of $I_\mathrm{r}$, typically a couple of error bars above the threshold value $R=1$.
 
\vspace*{-0.0cm}

\vspace*{-0.0cm}
 
\newpage
 
\subsection{Readout spectra}
\label{spec}

\begin{figure}[htb]
  \hspace{3.0cm}\includegraphics[width=10cm]{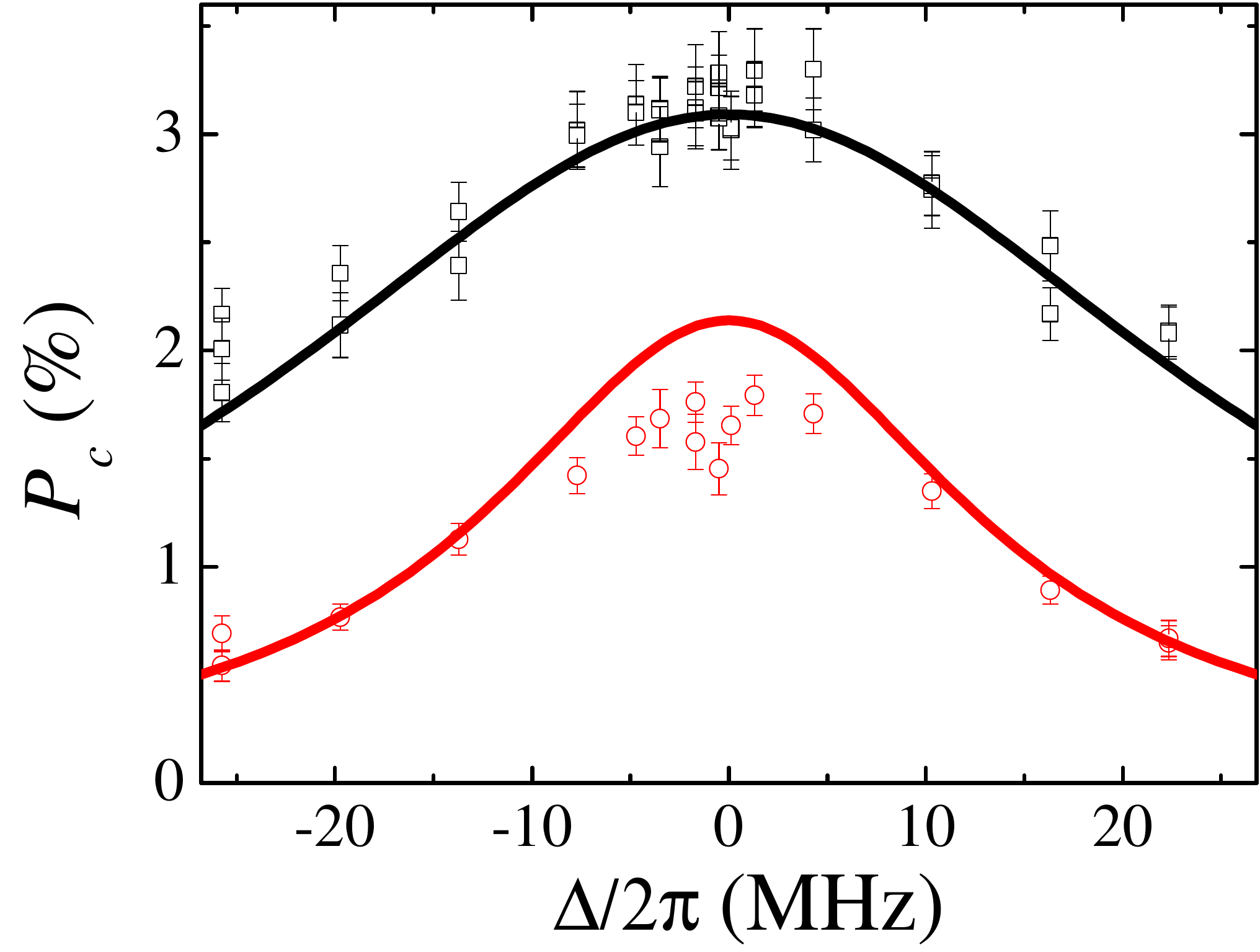}
  \vspace{-0.0cm}
  \caption{Total conditional probability $P_\mathrm{c}$ as a function of read-field detuning $\Delta$ for two intensities: $I_\mathrm{r} = 127$~mW/cm$^2$ (black squares) and 24~mW/cm$^2$ (red circles). The solid lines are the corresponding theoretical results obtained from (\ref{Pc-theory}). The parameters used are the same of figure ~\ref{fig2}, with the exception of the now $F=4.8$.}\label{fig7}
\end{figure}

\begin{figure}[htb]
  \hspace{3.0cm}\includegraphics[width=10cm]{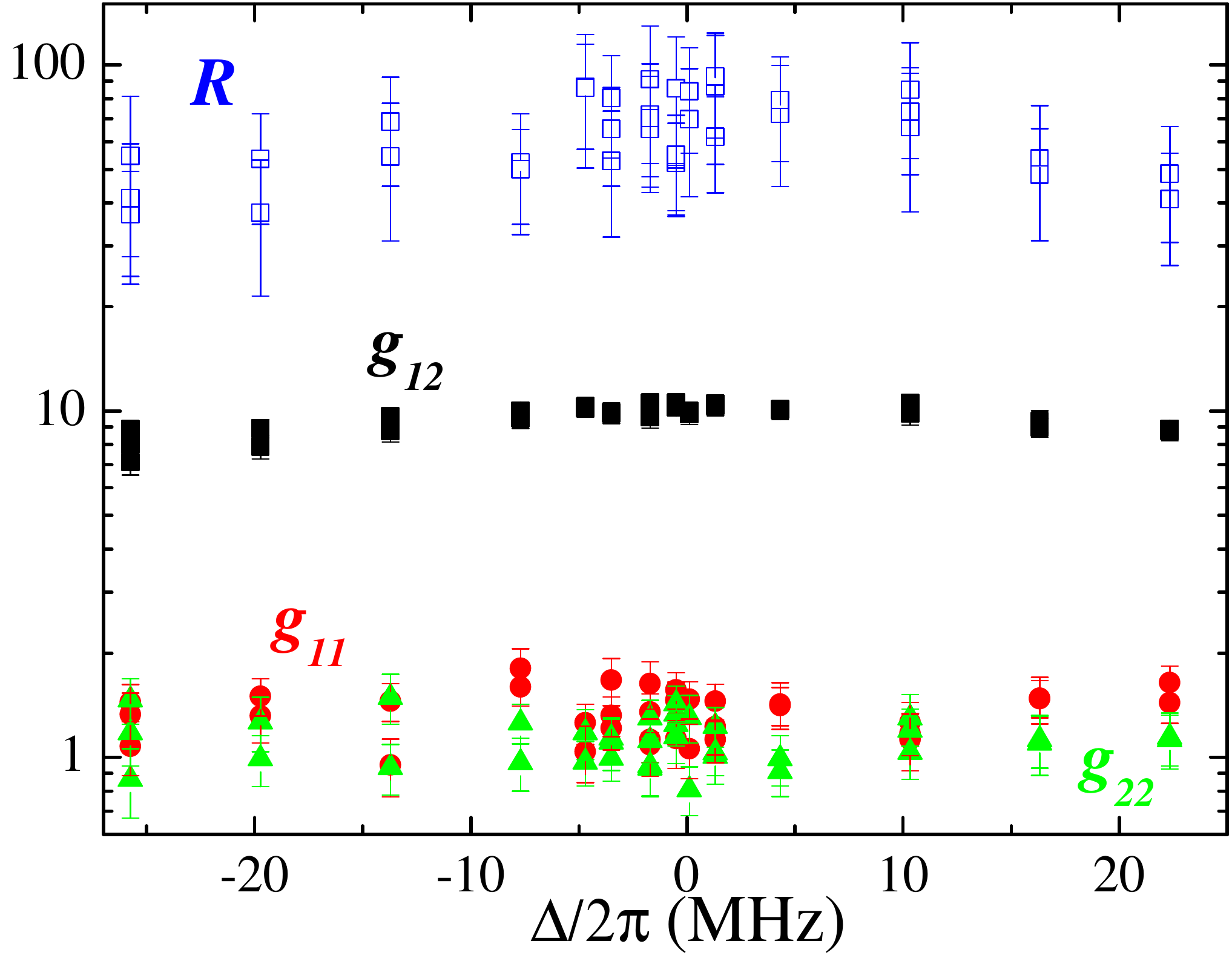}
  \vspace{-0.0cm}
  \caption{Experimental quantities characterizing the correlation between fields 1 and 2 as a function of the read detuning $\Delta$, for $I_\mathrm{r} = 127$~mW/cm$^2$. The black, solid squares are the results for the normalized cross-correlation function between field 1 and 2, $g_{12}$. The red, solid circles (green, solid triangles) are the results for the normalized auto-correlation function of field 1 (2), $g_{11}$ ($g_{22}$). The blue, open squares are the results for function $R$, which indicates the nonclassical nature of the correlations if $R>1$.}\label{fig8}
\end{figure}

Figure~\ref{fig7} plot integrated values of $P_\mathrm{c}$, such as in figure ~\ref{fig6}, but now as a function of the read-field detuning $\Delta$, for both $I_\mathrm{r} = 24$~mW/cm$^2$ and~127~mW/cm$^2$. This corresponds to measure the readout spectra of the quantum memory, as discussed in \cite{deOliveira12}. Differently from \cite{deOliveira12}, however, we are now in the regime of very strong read fields, which leads to the transparency of the medium to the extracted photon and to an enhanced probability of extraction at resonance ($\Delta = 0$). The corresponding theoretical curves are given by the solid lines. As discussed above, we employed here the same fitting parameters for the theory as in the previous figures, with exception of the value for $F$, which is now $F=4.8$, reflecting an optimized alignment for the field 2 detection at the time these plots were taken.  Figure~\ref{fig8} demonstrates the nonclassical nature of the stored state throghout the curve with $I_\mathrm{r} = 127$~mW/cm$^2$ in figure ~\ref{fig7}, in the same way as figure ~\ref{fig6}.
 
\vspace*{-0.0cm}
 
Such readout-spectra measurements reveal a systematic deviation from the theory at resonance for lower reading intensities, as can be seen in the results for $I_\mathrm{r} = 24$~mW/cm$^2$ in figure ~\ref{fig7}. This is expected, since our theory does not take the propagation of field 2 into account. We assume a strong transparency for the extracted photon. As discussed in detail in \cite{deOliveira12}, once the read intensity decreases the system becomes less transparent to field 2, which starts to be more absorved at resonance. On the other hand, we continue to observe a good quantitative agreement between theory and experiment on the wings of the spectra, even for low $I_\mathrm{r}$, since this region is less affected by the propagation and reabsorption of field 2.

\vspace*{-0.0cm}
 
\section{Conclusion}
\label{sec:conclusion}

We reported an investigation of the reading process of a quantum memory consisting of a collective state of an ensemble of cold atoms holding a single excitation. The reading process maps the stored excitation into a single photon, whose time dependency reveals then the dynamics of the reading process itself. We performed a series of experiments varying both intensity and detuning of the reading field, obtaining the wavepacket of the extracted photon in a variety of situations. Our experimental results were then employed to corroborate a simplified model for the reading process, which leads to particularly simple analytical expressions for the photonic wavepacket. The quantitative agreement observed between theory and experiments indicates then that we were able to capture the essential physical aspects of the problem in our model. This theoretical model also highlights and clarifies the role of superradiance in the system, describing how it affects not only the efficiency of the reading process, but also its saturation and spectrum. Further investigations are on the way to obtain more direct experimental measures of the effects related to superradiance, which is crucial to obtain the degree of efficiency in the reading process required by applications in the field of quantum information.

\section*{Acknowledgments}

We gratefully acknowledge Rafael de Oliveira for experimental assistance in various parts of this work and Marcos Aurelio for his technical assistance in the setup of the experiment. This work was supported by CNPq, CAPES, and FACEPE (Brazilian agencies), particularly through the programs PRONEX and INCT-IQ (Instituto Nacional de Ci\^encia e Tecnologia
de Informa{\c c}\~ao Qu\^antica).


\end{document}